\documentclass{ws-procs9x6-cpt22}
\begin{document}

\newcommand{\refeq}[1]{(\ref{#1})}
\def\etal {{\it et al.}}

\title{Lorentz symmetry breaking and supersymmetry}

\author{J. R. Nascimento, A. Yu. Petrov}

\address{$^1$Departamento de Física, Universidade Federal da Paraíba,\\
 Caixa Postal 5008, 58051-970, João Pessoa - PB, Brazil}

\begin{abstract}
We discuss three manners to implement Lorentz symmetry breaking in a superfield theory formulated within the superfield formalism, that is, deformation of the supersymmetry algebra, introducing of an extra superfield whose components can depend on Lorentz-violating (LV) vectors (tensors), and adding of new terms proportional to LV vectors (tensors) to the superfield action. We illustrate these methodologies with examples of quantum calculations.
\end{abstract}

\bodymatter

\section{Introduction}

The supersymmetry (SUSY) is treated now as an important ingredient of an expected unified model of fundamental interactions. Therefore, the natural question consists in a possibility of constructing supersymmetric extensions for known LV models. To do this, it is very convenient to use the superfield formalism known to be very advantageous (see f.e. Ref. \refcite{BK0}). To construct a LV superfield theory, one can introduce the Lorentz symmetry breaking at three possible levels -- (i) at the level of the SUSY algebra, (ii) at the level of the superfield, (iii) at the level of the superfield Lagrangian.

\section{Deformation of the SUSY algebra}

The most convenient method to construct superfield LV models is based on using the Kostelecky-Berger construction \cite{KB}. Within it, the deformed SUSY generators, in the four-dimensional spacetime, are defined as
\begin{equation}
Q_{\alpha}=i(\partial_{\alpha}-i\bar{\theta}^{\dot{\beta}}\sigma^m_{\dot{\beta}\alpha}\nabla_m);\quad\,
\bar{Q}_{\dot{\alpha}}=i(\partial_{\dot{\alpha}}-i\theta^{\beta}\bar{\sigma}^m_{\beta\dot{\alpha}}\nabla_m),
\end{equation}
where $\nabla_m=\partial_m+k_{mn}\partial^n$ is the "twisted" derivative, with $k_{mn}$ is a constant tensor implementing the Lorentz symmetry breaking. The analogous manner of modification of SUSY generators, i.e. replacement of $\partial_m$ by $\nabla_m$, is employed also within other representations of the SUSY algebra, as well as in the three-dimensional space-time. The $Q_{\alpha},\bar{Q}_{\dot{\alpha}}$ satisfy the deformed anticommutation relation: $\{Q_{\alpha},\bar{Q}_{\dot{\alpha}}\}=-2i\sigma^m_{\dot{\alpha}\alpha}\nabla_m$, while other anticommutators are not changed. The same deformation must be performed in spinor supercovariant dervatives $D_{\alpha},\bar{D}_{\dot{\alpha}}$  as well, to ensure their anticommuting with the SUSY generators. The $D_{\alpha},\bar{D}_{\dot{\alpha}}$ satisfy the relations:
\begin{eqnarray}
&& \{D_{\alpha},\bar{D}_{\dot{\alpha}}\}=2i\sigma^m_{\dot{\alpha}\alpha}\nabla_m;\quad\,
D^2\bar{D}^2D^2=16\tilde{\Box}D^2;\quad\,
D_{\alpha}D_{\beta}D_{\gamma}=0.
\end{eqnarray}

The superfield actions defined in a superspace with the deformed SUSY algebra formally reproduce the same expressions as in usual superfield theories, but being rewritten in components, they involve additional LV terms, f.e. the LV Wess-Zumino (WZ) model in components looks like
\begin{eqnarray}
S&=&\int d^4x \Big[\phi \tilde{\Box}\bar{\phi}+\psi^{\alpha}i\sigma^m_{\alpha\dot{\alpha}}\nabla_m\bar{\psi}^{\dot{\alpha}}+F\bar{F}+\nonumber\\&+&
\Big(m(\psi^{\alpha}\psi_{\alpha}+\phi F)+\lambda(\phi\psi^{\alpha}\psi_{\alpha}+\frac{1}{2}\phi^2 F)+h.c.\Big)
\Big],
\end{eqnarray}
i.e. it involves CPT-even aether-like terms \cite{Carroll} for scalar and spinor fields. Here, $\tilde{\Box}=\nabla^m\nabla_m$ is the deformed d'Alembertian operator.

The superfield propagators in this theory look like
\begin{eqnarray}
<\Phi(z_1)\bar{\Phi}(z_2)>=\frac{1}{\tilde{\Box}-m^2}\delta_{12};\quad\,
<\Phi(z_1)\Phi(z_2)>=\frac{mD^2}{4\tilde{\Box}(\tilde{\Box}-m^2)}\delta_{12}.
\end{eqnarray}
Their difference from the usual case consists in the presence of $\tilde{\Box}$. 

In this theory, one can calculate the one-loop low-energy effective action described by the K\"{a}hlerian effective potential. It is contributed by the series of supergraphs depicted at Fig. 1, with external legs are for alternating $\Psi=m+\lambda\Phi$ and $\bar{\Psi}=m+\lambda\bar{\Phi}$.

\begin{figure}
\includegraphics[width=4in]{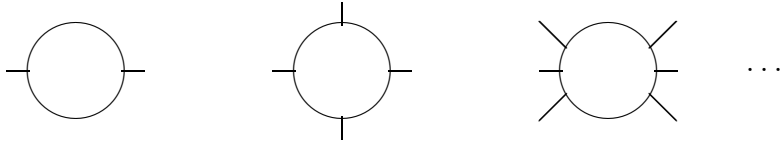}
\caption{One-loop contributions to the K\"{a}hlerian effetive potential.}
\label{aba:fig1}
\end{figure}

The sum of these supergraphs is easily found to be \cite{ourWZ}
\begin{eqnarray}
K^{(1)}=-\frac{1}{2}\int d^8z\int\frac{d^4q}{(2\pi)^4}\frac{1}{(q_m+k_{mn}q^n)^2}\ln(1-\frac{\bar{\Psi}\Psi}{(q_m+k_{mn}q^n)^2}).
\end{eqnarray}
We change $q_m+k_{mn}q^n\to \tilde{q}_m$:
\begin{eqnarray}
K^{(1)}=-\frac{1}{2}\Delta\int d^8z\int\frac{d^4\tilde{q}}{(2\pi)^4}\frac{1}{\tilde{q}^2}\ln(1-\frac{\bar{\Psi}\Psi}{\tilde{q}^2}),
\end{eqnarray}
where $\Delta=\det(\frac{\partial q^m}{\partial\tilde{q}^n})=\det^{-1}(\delta^m_n+k^m_n)$ is a Jacobian of change of variables. After integration and renormalization we arrive at
\begin{eqnarray}
K^{(1)}=-\frac{1}{32\pi^2}\Delta\int d^8z \Psi\bar{\Psi}\ln\frac{\Psi\bar{\Psi}}{\mu^2}.
\end{eqnarray}
We conclude that in this case quantum calculations are no more difficult that in the usual Lorentz-invariant case. This approach was also  generalized for supergauge theories \cite{ourQED}.

\section{Introduction of the additional superfield}

Within this approach we introduce a new superfield whose some components depend on LV vectors (tensors). The most interesting example of its application allowed to construct the SUSY extension of the Carroll-Field-Jackiw (CFJ) term \cite{Hel}: we introduce the new superfield $S=s+\ldots$, where $s$ is its lower component, and define the CPT-odd term
\begin{equation}
S_{odd}=\int d^8z S W^{\alpha}D_{\alpha}V+h.c., 
\end{equation}
where $V$ is the gauge scalar superfield, and $W_{\alpha}=-\frac{1}{4}\bar{D}^2D_{\alpha}V$ is the corresponding superfield strength. In components this term looks like
\begin{equation}
S_{odd}=\frac{i}{2}\int d^4x\partial_a(s-s^*)\epsilon^{abcd}F_{bc}A_d+\ldots,
\end{equation}
thus, if $s(x)=-ik_ax^a$, we reproduce the CFJ term. Actually, this is the only known manner to obtain the CFJ term within the superfield framework. This methodology can be used to obtain a superfield extension for the CPT-even aether-like term as well \cite{Hel2}.

\section{Direct modification of the superfield Lagrangian}

In this case, we add Lorentz-breaking term to a Lagrangian. The simplest example is the following Horava-Lifshitz-like action of the chiral superfield, with $z$ is the critical exponent measuring the space-time anisotropy:
\begin{equation}
\label{sfigeral}
S=\int d^8z\Phi(1+\rho\Delta^{z-1})\bar{\Phi}+(\int d^6z W(\Phi) +h.c.),
\end{equation}
with $W(\Phi)$ is a some potential. For this theory, we can calculate the one-loop low-energy effective action, using the same scheme as in the section 2, with $\Psi=W^{\prime\prime}$, the supergraphs are again given by Fig. 1, and the propagators are correspondingly modified. As a result, we arrive at
\begin{equation}
\label{k1min}
K^{(1)}=-\frac{i}{2}\int d^4\theta\frac{d^4k}{(2\pi)^4}\frac{1}{k^2}\ln\left[1+\frac{\Psi\bar{\Psi}}{k^2(1+\rho(\vec{k}^2)^{z-1})^2}
\right].
\end{equation}
This integral can be calculated only approximately. Under the scheme developed in \cite{HLSUSY}, where subleading orders of $\vec{k}$ are disregarded, we find
\begin{equation}
\label{epmin}
K^{(1)}=\frac{1}{12\pi}\csc(\frac{\pi}{z})
(\frac{4\rho}{3})^{-1/z} (\Psi\bar{\Psi})^{1/z}.
\end{equation}
This expression is singular at $z=1$, as it must be, since this choice of $z$ is necessary to achieve the Lorentz invariance.

\section{Summary}

We presented three schemes for constructing LV superfield theories. It turns out to be that, first, in principle, perturbative calculations in LV superfield theories are no more difficult as in usual ones, second, the CPT-even scenario is easier to reproduce in a superfield form. So, the main problems to study are, first, search for other manners to construct LV superfield theories, second, generalization of these schemes to a curved space-time.

\section*{Acknowledgments}
This study was partially supported by Conselho Nacional de Desenvolvimento Cient\'\i fico e Tecnol\'ogico (CNPq) via the grant 301562/2019-9.

\end{document}